\documentclass[times,twocolumn,final]{elsarticle}

\usepackage{framed,multirow}

\usepackage{amssymb}
\usepackage{latexsym}

\usepackage{url}
\usepackage{xcolor}
\usepackage{hyperref}

\definecolor{newcolor}{rgb}{.8,.349,.1}

\begin{document}

\begin{frontmatter}
\title{Attri-VAE: attribute-based interpretable representations of medical images with variational autoencoders}

\author[1]{Irem {Cetin}}
\author[1,2]{Maialen {Stephens}}
\author[1]{Oscar {Camara}}
\author[1,3]{Miguel A. {Gonz\'alez Ballester}\corref{cor1}}
\cortext[cor1]{Corresponding author: }
\ead{ma.gonzalez@upf.edu}


\address[1]{BCN Medtech, Department of Information and Communication Technologies, Universitat Pompeu Fabra, Barcelona, Spain }
\address[2]{Vicomtech Foundation, San Sebasti\'an, Spain }
\address[3]{ICREA, Barcelona, Spain }


\begin{abstract}

\textit{Deep learning (DL) methods where interpretability is intrinsically considered as part of the model are required to better understand the relationship of clinical and imaging-based attributes with DL outcomes, thus facilitating their use in the reasoning behind the medical decisions. Latent space representations built with variational autoencoders (VAE) do not ensure individual control of data attributes. Attribute-based methods enforcing attribute disentanglement have been proposed in the literature for classical computer vision tasks in benchmark data. In this paper, we propose a VAE approach, the Attri-VAE, that includes an attribute regularization term to associate clinical and medical imaging attributes with different regularized dimensions in the generated latent space, enabling a better-disentangled interpretation of the attributes. Furthermore, the generated attention maps explained the attribute encoding in the regularized latent space dimensions. Using the Attri-VAE approach we analyzed healthy and myocardial infarction patients with clinical, cardiac morphology, and radiomics attributes. The proposed model provided an excellent trade-off between reconstruction fidelity, disentanglement, and interpretability, outperforming state-of-the-art VAE approaches according to several quantitative metrics. The resulting latent space allowed the generation of realistic synthetic data in the trajectory between two distinct input samples or along a specific attribute dimension to better interpret changes between different cardiac conditions.}

\end{abstract}


\end{frontmatter}



\section{Introduction}

\label{introduction}
Deep learning (DL) methods have recently shown great success in many fields, from computer vision \citep{DLinCV, 8237506, goodfellow_generative_2014} to natural language processing \citep{ DLinNLP1,  DLinNLP}, among numerous others. In addition, DL methods have started to dominate the medical imaging field \citep{DLinMedImg}, being used in a variety of medical imaging problems, such as segmentation of anatomical structures in the images \citep{bernardTMI, UNET, LpezLinares2018FullyAD}, disease prediction \citep{pmid31481890}, medical image reconstruction \citep{pmid31818389, exrecons2020} and clinical decision support \citep{sergio19}. Despite achieving exceptional results, DL methods face challenges when applied to medical data regarding explainability, interpretability, and reliability because of their underlying black-box nature \citep{jimaging6060052, 2021210031}. Hence, the need for tools that investigate interpretability in DL is also emerging in healthcare.

Recent reviews of interpretable DL can be found in \citep{jimaging6060052, BARREDOARRIETA202082, molnar2019, serg2021}. Some methods have been proposed that employ backpropagation-based attention maps to either generate class activation maps that visualize the regions with high activations in specific units of the network \citep{gradcam} or saliency maps using gradients of the inputs with respect to the outputs \citep{Simonyan2014DeepIC, Kapishnikov2019XRAIBA}. Other methods also proposed creating proxy models that focus on complexity reduction such as LIME \citep{LIME_Ribeiro0G16} or by approximating a value based on game theory optimal Shapley values to explain the individual predictions of a model \citep{NIPS2017_8a20a862}. However, it is key to design models that are inherently interpretable, rather than creating posthoc models to explain the black-box ones \citep{Rudin2019}. 

Recently, models based on latent representations, such as variational autoencoders (VAE), have become powerful tools in this direction \citep{Liu2020TowardsVE, Biffi2020ExplainableAS}, as their latent space is able to encode important hidden variables of the input data \citep{Kingma2014AutoEncodingVB}. Especially, when dealing with data that contains different interpretable features (data attributes), it is interesting to see how and if these attributes have been encoded in the latent space. Even though the proposed approaches provide promising results, they have some limitations, one of which is that the encoded variables cannot be easily controlled; they mostly show an entangled behavior, meaning each latent factor maps to more than one aspect in the generative process \citep{PMID23787338}.

In order to bypass this limitation, much effort has been done to enforce disentanglement in the latent space \citep{Higgins2017betaVAELB, factorvae, Rubenstein2018LearningDR, MIG_metric,CHARTSIAS2019101535}, being the majority of them unsupervised techniques \citep{PMID23787338, Locatello2019ChallengingCA}. While many of these methods show good disentanglement performance, they are not only sensitive to inductive biases (e.g., choice of the network, hyperparameters, or random seeds), but also some amount of supervision is necessary for learning effective disentanglement \citep{Locatello2019ChallengingCA}. Moreover, since these methods are able to learn a factorized latent representation without attribute specification, they require a posthoc analysis to determine how different attributes are encoded to different dimensions of the latent space \citep{arvae}. 

On the other hand, attribute-based methods aim to establish a correspondence between data attributes of interest and the latent space \citep{Hadjeres2017GLSRVAEGL, lample2017fader, Bouchacourt2018MultiLevelVA, arvae}. However, these methods also have their drawbacks: some of them are limited to work only on certain types of data attributes \citep{lample2017fader}; some impose additional constraints \citep{Bouchacourt2018MultiLevelVA}; very few of them are designed to work with continuous variables \citep{Hadjeres2017GLSRVAEGL, arvae}; some require differentiable computation of the attributes; and they are extremely sensitive to the hyperparameters \citep{Hadjeres2017GLSRVAEGL}. However, \citep{arvae} have recently shown promising results for interpretability with their approach, associating each data attribute to a different regularized dimension of the latent space, which they have applied in the MNIST database for digit number recognition. The same approach was also employed as a post-processing step to generate interpretable and temporally consistent segmentations of echocardiography images \citep{echoARVAE}.

In this paper, we proposed a VAE-based approach (Attri-VAE), as it is able to encode certain hidden attributes of the data \citep{carter2017using} which can then be used to control data generation, and thus, improve the interpretation of the data attributes \citep{arvae}. Attri-VAE is an attribute-interpreter VAE based on attribute-based regularization \citep{arvae} in the latent space, for an enhanced interpretation of clinical and imaging attributes obtained from multi-modal sources. Additionally, we also employed a classification network (MLP) that enables to identify different clinical conditions, e.g., healthy vs. pathological cases. Furthermore, we incorporate gradient-based attention map computation \citep{Liu2020TowardsVE} to visually explain our proposed network by generating attention maps that show the high-response regions for each value of data attributes that are encoded in the regularized latent space dimensions. The main contributions of this work can be described as follows:
\begin{itemize}

    \item The proposed approach is able to interpret different data attributes where specific ones are forced to be encoded along specific latent dimensions without the need for any posthoc analysis, while encouraging attribute disentanglement by employing $\beta$-VAE as a backbone \citep{Higgins2017betaVAELB}.
    
    \item The structured latent space enables controllable data generation by changing the latent code of the regularized dimension (i.e., following the corresponding attribute), generating new data samples as a result of manipulating these dimensions. For instance, if the attribute represents the volume in a region of interest (ROI) and the corresponding regularized dimension is the first one of the latent code, then increasing values of this dimension would result in an increase in the ROI volume. The ability to generate a heterogeneous set of medical images is promising as collecting large annotated images is especially an issue in the medical image domain. As our approach allows the generation of diverse datasets, it may be applied in a variety of clinical scenarios. It could, for example, be used to train robust deep-learning models, or it may be trained with various clinical conditions to generate synthetic images by modifying the latent code of regularized dimensions that follows the corresponding data attribute. Furthermore, a controlled virtual cohort generation could also be employed to remove the burden of needing to gather and select real data for clinical trials.
    
    \item Attribute-based gradient-based attention maps provide a way to explain how the gradient information of individual attributes flow inside the proposed architecture by showing high-response regions in the images.
   
    \item The classification network provides a way to stratify different cohorts, based on the attributes in the latent space. In this way, the most discriminative features for the classification task are identified by projecting original samples into the latent space.
     
\end{itemize}

\begin{figure*}[!t]
\centering
\includegraphics[width=\textwidth]{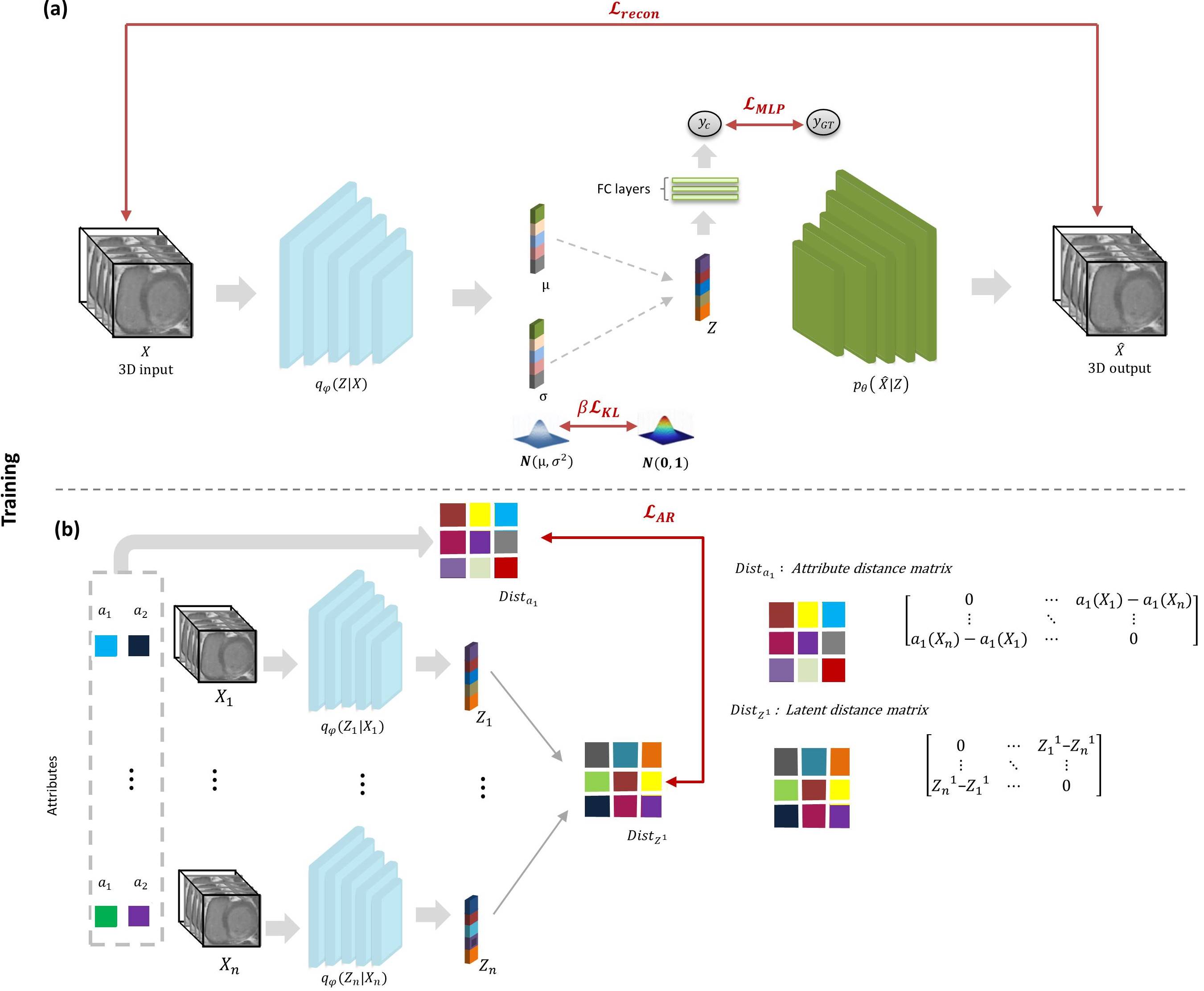}
\caption{Training framework of the proposed approach. Loss functions are shown in red arrows. The total loss function of the model is: $\mathcal{L}=\mathcal{L}_{recon}+ \beta \mathcal{L}_{KL} + \mathcal{L}_{MLP}+ \gamma \mathcal{L}_{AR}$.
(a) Losses computed for each data sample: multilayer perceptron (MLP) loss ($\mathcal{L}_{MLP}$), Kullback-Leibler (KL) loss ($\mathcal{L}_{KL}$), and reconstruction loss ($\mathcal{L}_{recon}$). (b) Attribute-regularization loss ($\mathcal{L}_{AR}$), computed inside a training batch that has $n$ data samples. The input, a 3D image ($X$), first goes through the 3D convolutional encoder, $q_\varphi (Z \vert X)$, which learns to map $X$ to the low dimensional space $Z$ by outputting the mean ($\mu$) and variance ($\sigma$) of the latent space distributions. The decoder, $p_\theta (\hat{X} \vert Z)$,  then takes $Z$ and outputs the reconstruction of the original input, ($ \hat{X}$). The predicted classes of the inputs, $y_c$, are computed with a MLP module that consists of three fully connected (FC) layers. The corresponding MLP loss function is computed between $y_c$ and the ground truth label $y_{GT}$. In (b), $\mathcal{L}_{AR}$ is shown to regularize the first dimension of the latent space ($Z^1$) with the attribute $a_1$ ($a_1$ and $a_2$ represent the first and the second attributes, respectively). $Dist_{Z^1}$ is the distance matrix of the first latent dimension, while $Dist_{{a}_1}$ represents the distance matrix of the attribute $a_1$.}
\label{fig:training}
\end{figure*}

In this work, we have applied the proposed Attri-VAE approach to study cardiovascular pathological conditions, such as myocardial infarction, using the EMIDEC\footnote{http://emidec.com/} cardiac imaging dataset \citep{emidec}, including clinical and imaging features, also exploring the association with radiomics descriptors. Additionally, we used ACDC MICCAI17 database\footnote{https://www.creatis.insa-lyon.fr/Challenge/acdc/} as an external testing dataset.

The remainder of this paper is organized as follows. Firstly, we present the methodology and the details of our architecture in Section \ref{Methods}. We then describe the experimental setup and employed dataset in Section \ref{experimental_setup}. Section \ref{results} provides the results that are discussed in Section \ref{discussion}. Finally, in Section \ref{conclusions} we conclude our findings.

\section{Methodology}
\label{Methods}

\begin{figure*}[!t]
\centering
\includegraphics[width=\textwidth]{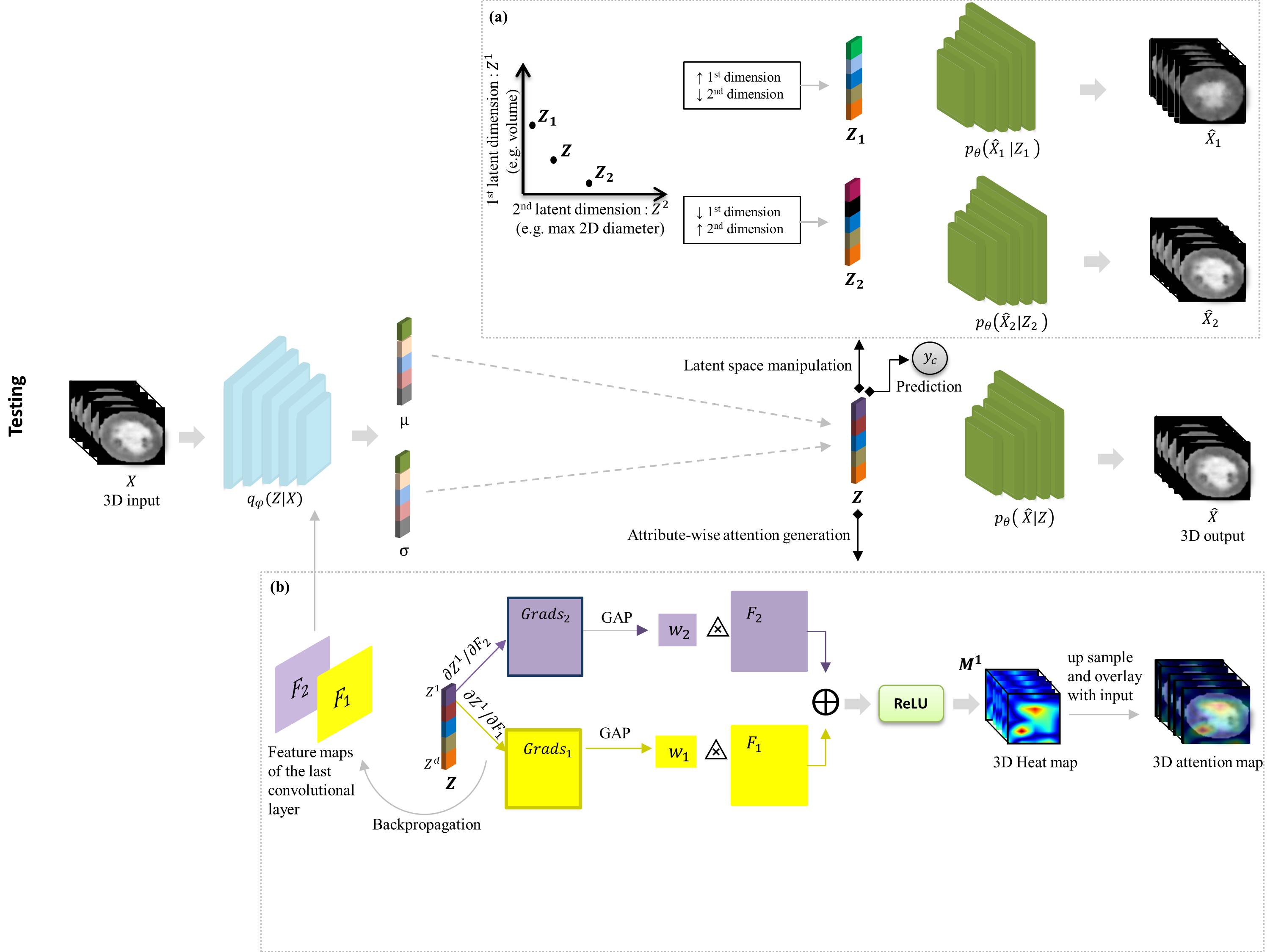}
\caption{The trained network can be used for: (a) latent space manipulation; and (b) generating attribute-based attention maps. For a given 3D data sample, $X$, the trained 3D convolutional encoder, $q_\varphi (Z \vert X)$, outputs the mean ($\mu$) and variance ($\sigma$) vectors, then $Z$ being sampled with the reparameterization trick. (a) Data generation process by changing only first ($Z^1$) and second ($Z^2$) regularized latent dimensions of $Z$, which correspond to two different data attributes (volume and maximum 2D diameter, respectively). Then, the decoder, $p_\theta (X \vert Z)$, generates 3D outputs, $X_1$ and $X_2$, using the manipulated latent vectors, $Z_1$ and $Z_2$, respectively. (b) Attribute-based attention map generation for a given attribute, which is encoded in the first latent dimension $(Z^1)$. First, $(Z^1)$ is backpropagated to the encoder´s last convolutional layer to obtain the gradient maps ($Grads_1$ and $Grads_2$) with respect to the feature maps ($F_1$ and $F_2$). The gradient maps of ($Z^1$) measure the linear effect of each pixel in the corresponding feature map on the latent values. After that, we compute the weights ($w_1$ and $w_2$) using global average pooling (GAP) on each gradient map. A heat map is generated by multiplying these values ($w_1, w_2$) with the corresponding feature map, summing them up and applying an activation unit (ReLU). Finally, the heat map is upsampled and overlaid with the input image to obtain the superimposed image (3D attention map). Additionally, the class score of the input, $y_c$, is computed with the multilayer perceptron (MLP) that is connected to $Z$. Note that, in the figure it is assumed that the last convolutional layer of the encoder has 2 feature maps.}
\label{fig:testing}
\end{figure*}

The overall structure of our framework is shown in Figure \ref{fig:training} (training) and Figure \ref{fig:testing} (testing). The proposed Attri-VAE incorporates attribute regularization into a $\beta$-VAE framework that was used as a backbone for the interpretation of data attributes. The trained network enables to generate new data samples by manipulating the data attributes, whereas the generated attribute-based attention maps explain how the gradient information of each attribute flows inside the proposed architecture.  
This section is organized firstly explaining the overall training criterion of the proposed model, with the following subsections describing each of the elements of our methodology and their integration.

\subsection{Training criterion}
\label{training_criterion_of_the_model}

Attri-VAE is trained with a loss function, $\mathcal{L}$, which is composed of four terms, as follows:  \\

\begin{equation}
\label{loss_all}
\mathcal{L}=\mathcal{L}_{recon}+ \beta \mathcal{L}_{KL} + \mathcal{L}_{MLP}+ \gamma \mathcal{L}_{AR}.
\end{equation}

The reconstruction loss,  $\mathcal{L}_{recon}$, is based on the binary cross-entropy (BCE) between the input $X$ and its reconstruction $\hat{X}$, while the second term, $\mathcal{L}_{KL}$, employs the Kullback-Leibler (KL) divergence between the learned prior and the posterior distributions, weighted by a hyperparameter ($\beta$). An additional term, $\mathcal{L}_{MLP}$, estimates the BCE loss for the classification between the network prediction, $y_c$, and the ground truth label, $y_{GT}$. The final loss term,  $\mathcal{L}_{AR}$, includes the attribute regularization, with a tunable hyperparameter ($\gamma$) that weights its strength. In the following sections, detailed explanations of each loss term in our training criterion can be found (also see Figure \ref{fig:training}). 

\subsection{ Variational autoencoder (VAE) and $\beta$-VAE}
\label{VAE}

A variational autoencoder \citep{Kingma2014AutoEncodingVB} is a generative model that consists of an encoder and a decoder, and aims to maximize the marginal likelihood of the reconstructed output, which is written as:\\



\begin{equation}
log_{p_\theta } (X) \geq \mathbb{E}_{Z \sim q_\varphi (Z \vert X) } [log p_\theta (X \vert Z)] - D_{KL} (q_\varphi (Z \vert X) \Vert p(Z)) 
\end{equation}
In this objective function, the first term is the log likelihood expectation that the input $X$ can be generated by the sampled $Z$ from the inferred distribution, $q_\varphi (Z|X)$. The second term corresponds to the KL divergence between the distribution of $Z$ inferred from $X$, and the prior distribution of $Z$. Note that both distributions are assumed to follow a multivariate normal distribution. 

In practice, the loss function of the VAE consists of two terms: a first term that penalizes the reconstruction error between the input and output; and a second term forcing the learned distribution, $q_\varphi (Z|X)$, to be as similar as possible to the prior distribution, $p(Z)$. In this case, the overall VAE loss can be written as:\\

\begin{equation}
\mathcal{L}_{VAE} (\theta,\varphi)= \mathcal{L}_{recon} (\theta,\varphi)+ \mathcal{L}_{KL} (\theta,\varphi),
\end{equation}
where the reconstruction loss, $\mathcal{L}_{recon} (\theta,\varphi)$, and the KL loss, 
$\mathcal{L}_{KL} (\theta,\varphi)$, are computed as follows: \\

\begin{equation}
\label{eq:recon_loss}
\mathcal{L}_{recon} (\theta,\varphi) = \sum_{i=1}^N {\Vert  \hat{X} - X \Vert}_2^2,
\end{equation}
\begin{equation}
\label{eq:KL_loss}
\mathcal{L}_{KL} (\theta,\varphi) = \mathcal{D}_{KL} (q_\varphi (Z|X) \Vert p(Z)  ).
\end{equation} 



In this work we chose to use $\beta$-VAE as the backbone of our approach to encourage the disentanglement as it is easy to formulate and it has shown good performance based on one or more disentanglement metrics \citep{Higgins2017betaVAELB, burgess2018understanding}. 




The $\beta$-VAE approach \citep{Higgins2017betaVAELB} is an extension of the standard VAE that aims to learn a disentangled representation of the encoded variables in a completely unsupervised manner \citep{Locatello2019ChallengingCA, Higgins2017betaVAELB} by simply giving more weight to the KL term, compared to the original VAE, with an extra hyperparameter $\beta$:\\

\begin{equation}
\label{eq:betaVAE_loss}
\mathcal{L}_{VAE} (\theta,\varphi)= \mathcal{L}_{recon} (\theta,\varphi)+  \beta \mathcal{L}_{KL} (\theta,\varphi),
\end{equation}


\subsection{Attribute-based regularization}
\label{attribute_reg}
%

In order to better interpret the data attributes that are encoded in the latent space, we employ an attribute-based regularization loss \citep{arvae}, which aims at encoding encode an attribute $a$ along a dimension $d$ of the latent space (regularized dimension).

The attribute regularization loss, $\mathcal{L}_{AR}$, is calculated for the dimension $d$ of the latent space in a training batch containing $n$ training examples for the purpose of forcing the dimension $d$ to have a monotonic relationship with the attribute values of $a$. The attribute regularization loss is then computed as follows: 
\begin{equation}
\label{eq:ar}
\mathcal{L}_{AR}(d,a)= MAE ( tanh ( \delta {Dist}_{Z^d} ) - sgn( {Dist}_{a})), 
\end{equation}
where $MAE$ is the mean absolute error, ${Dist}_{a}$ is the attribute distance matrix, and ${Dist}_{Z^d}$ is the distance matrix of the latent dimension $d$. These matrices are computed for all $n$ data examples in the corresponding training batch, such that: 

\begin{equation}
    \label{eq:dist_r}
    {Dist}_{a}=a(X_i )- a(X_j ),
\end{equation}
\begin{equation}
    \label{eq:dist_d}
    {Dist}_{Z^d}= Z_i^d- Z_j^d ,
\end{equation}
where $i, j \in [0, n)$, $X_i$ and $X_j$ are two exemplary samples (Equation \ref{eq:dist_r}), and each $D$-dimensional latent vector is represented as $Z =  \{Z^d\} $, where $d \in [0, D)$ (Equation \ref{eq:dist_d}).

In Equation \ref{eq:ar}, $tanh$ and $sgn$ refer to hyperbolic tangent function and sign function, respectively, whereas $\delta$ is the hyperparameter that modulates the spread of the posterior distribution.
For multiple selected attributes of interest to be encoded in the latent space, the overall loss function can be computed by summing all the corresponding objective functions together. Specifically, when the attribute set is $A: \{a_k\}$, where $k \in [0,K)$ contains $K$ attributes ($K \leq D$, being $D$ the latent size), then the overall loss function is computed as: 
\begin{equation}
\label{eq:AR_total_loss}
\mathcal{L}_{AR}=\sum_{k=0}^{K-1} \mathcal{L}_{d_k,a_k },
\end{equation}
where $d_k$ represents the index of the regularized dimension for the attribute $k$. This process is represented in Figure \ref{fig:training} (b).

\subsection{Classification network}
\label{mlp}

Recently, performing a classification task using VAEs has been proposed to learn and separate different cohorts in the latent space. For example, Biffi et. al. \citep{Biffi2020ExplainableAS} classified heart pathologies with cardiac remodelling using  explainable task-specific shape descriptors learned directly with a VAE architecture from the input segmentations. Additionally, other approaches based on VAE have also been applied to analyse coronary artery diseases \citep{clough}, Alzheimer's disease \citep{vae_mlp_alzheimer} or to predict the response of cardiomyopathy patients to cardiac resynchronization therapy \citep{pmid34109325}.

In this line, to enforce class separation to the Attri-VAE, a multilayer perceptron (MLP) prediction network was connected to the latent vector,  $p(y_c \vert Z)$ ( see Figure \ref{fig:training}). The corresponding objective function can be computed as the binary cross entropy (BCE) between the network prediction $y_c$ and the ground truth label $y_{GT}$, such that:

\begin{equation}
\label{eq:MLP_loss}
 \mathcal{L}_{MLP} = BCE(y_c, y_{GT})
\end{equation}

\subsection{Attribute-based attention generation}
\label{attention_generation}
The Attri-VAE facilitates data interpretation by generating new data samples as a result of scanning the regularized latent dimensions. Furthermore, it also provides a way to obtain attention maps from these dimensions (attribute-based attention map generation) for a better understanding on how gradient information of these attributes flows inside the proposed architecture (as can be seen in Figure \ref{fig:testing}).

Attribute-based visual attention maps were generated by means of gradient-based computation (Grad-CAM) \citep{gradcam}, as proposed by \citep{Liu2020TowardsVE}. Basically, a score is calculated from the latent space that is then used to estimate the gradients and attention maps. Specifically, given the posterior distribution inferred by the trained network for a data sample $X$, $q_\varphi (Z|X)$, the corresponding $D$-dimensional latent vector $Z$ is sampled using the reparameterization trick \citep{Kingma2014AutoEncodingVB}. Subsequently, for a given attribute set $A: \{a_k\}$, where $k \in [0,K)$ contains $K$ attributes, attribute-based attention maps, $M^{d_k}$, are generated for each regularized latent dimension $Z^{d_k}$ by backpropagating the gradients to the encoder´s last convolutional feature maps ($F:  \{F_i\}$ where $i \in [0,n)$):

\begin{equation}
    M^{d_k} = ReLU(\sum_{i=1}^n w_i F_i),
\end{equation}

\noindent where $d_k$ is index of the regularized latent dimension for a given attribute $k$. The weights, $w_i$, are computed using global average pooling (GAP), which allows us to obtain a scalar value, as follows:

\begin{equation}
    w_i = GAP( \frac {\partial Z^{d_k}}{\partial F_i})
              = \frac{1}{T} \sum_{p=1}^j \sum_{q=1}^l (\frac{\partial Z^{d_k}}{\partial F_i^{pq}}),
\end{equation}
where $T= j \times l$, (i.e., $ width \times height$), and $F_i^{pq}$ is the pixel value at location $(p,q)$ of the $j \times l$ matrix $F_i$. This process is visually summarized in Figure \ref{fig:testing}.

\section{Application for interpretable cardiology}
\label{experimental_setup}


\subsection{Datasets}
\label{sec:Dataset}
Initially, the EMIDEC dataset \citep{emidec} was used in our experiments. It is a publicly available database with delay-enhancement magnetic resonance images (DE-MRI) of 150 cases (100 and 50 cases for training and testing, respectively), with the corresponding clinical information. Each case includes a DE-MRI acquisition of the left ventricle (LV), covering from base to apex. The training set, with ground-truth segmentations, includes 67 myocardial infarction (MINF) cases and 33 healthy subjects. The testing set includes 33 MINF and 17 healthy subjects. Some clinical parameters were also provided along with the MRI: sex, age, tobacco (yes, no, and former), overweight, arterial hypertension, diabetes, family history of coronary artery disease, electrocardiogram (ECG), Killip max\footnote{A score based on physical examination and the development of the heart failure to predict the risk of mortality.}, troponin\footnote{A parameter that shows the level of the protein that is released into the bloodstream.}, LV ejection fraction (EF), and NTproBNP\footnote{A parameter that shows a level of a peptide, which is an indicator for the diagnosis of heart failure.}. Furthermore, we also used an additional external testing dataset for a more robust assessment of the classification performance, the ACDC MICCAI17 challenge training dataset \footnote{https://www.creatis.insa-lyon.fr/Challenge/acdc/}  (end-diastole, ED, and end-systole, ES, cine-MRI from 20 healthy volunteers and 20 MINF cases). The ACDC dataset includes ground-truth segmentations of the left ventricle, myocardium, and right ventricle by an experienced manual observer at both ED and ES time points \citep{bernardTMI}. The reader is referred to \citep{emidec,bernardTMI} for more details on the MRI acquisition protocol.

As a pre-processing step, the intensities of the left ventricle in all images were scaled between 0 and 1. Additionally, each image was cropped and padded ($x = 80$; $y = 80$; $z = 80$; $t = 1$).



\subsection{Cardiac attributes}
\label{sec:attributes}
Three different types of attributes were studied in our experiments. Initially, the Attri-VAE was trained with cardiac shape descriptors (e.g., wall thickness, LV and myocardial volumes, ejection fraction), extracted from ground-truth segmentations, which can easily be visually interpreted. In addition, attributes available in the clinical information with the highest discriminative performance were identified using recursive feature elimination (RFE) with a support vector machine (SVM) classification model (linear kernel, regularization parameter  $C=10$) since this approach has already shown good performance for feature selection tasks \citep{pmid25295306, Samb2012ANR, pmid29138828}. In total 12 clinical attributes were provided with the EMIDEC dataset as introduced in Section \ref{sec:Dataset} and the most discriminative attributes were then included in our analysis (e.g., gender, age, tobacco). The feature selection pipeline was done using the python-based machine learning library scikit-learn (version 1.0.2).\footnote{https://scikit-learn.org/stable/}

Finally, the Attri-VAE was also trained with radiomics features. Radiomics analysis was originally proposed to capture alterations at both the morphological and tissue levels in oncology applications\citep{Aerts2014DecodingTP, pmid28975929}, deriving multiple quantifiable features from pixel-level data. More recently, radiomics approaches have provided promising results on cardiac MRI data, for discriminating different cardiac conditions \citep{neisius, Larroza, Baessler, cetin2}, and to study cardiovascular risk factors in large databases \citep{cetin1}. Radiomics analysis represents a step towards interpretability compared to other black-box approaches since some features can be related to pathophysiological mechanisms \citep{cetin1}. However, there is a need for improving the robustness and reproducibility of radiomics outcomes across different feature selection strategies and imaging protocols, which would lead to enhanced explainability. For this reason, radiomics features were employed in our experiments to benefit from the proposed network's ability to explain the encoded attributes. The open-source library PyRadiomics (version 3.0.1) \footnote{https://pyradiomics.readthedocs.io/} was used to derive 214 features per analysed cardiac structure including 28 shape-based, 36 intensity-based, and 150 texture-based features. Subsequently, radiomics features with the highest discriminative performance were identified using the above-mentioned feature selection approach as this strategy has also demonstrated good performance in previous radiomics studies \cite{pmid28199039, pmid31799873, Chen2018ComputerAidedGO}. The top-performing features of this process were then selected to train the Attri-VAE. 

\subsection{Architectural details}
\label{sec:architecture_details}
The 3D convolutional encoder of the proposed Attri-VAE framework compresses the input into a 250-dimensional embedding through a series of 5 3D convolutional layers with kernel size 3 and stride 2, except the last convolutional layer that has stride 1. Note that, using a series of 3 fully connected layers, the latent dimension was set to 64.
The prediction network was constructed with a shallow 3-layer MLP to be able to discriminate between the healthy and infarct subjects, using a ReLU activation function as a non-linearity after the first two layers. The upsampling and convolutional layers used in the encoder and the decoder were followed by batch normalization and ReLU non-linearity, except the decoder's last convolutional layer (Attri-VAE output) where a sigmoid function was applied. All the network weights were randomly initialized with Xavier initialization \citep{xavier}. The tunable parameters of the loss function (Equation \ref{loss_all}) were fixed as follows: KL weight $\beta = 2$; and regularization weight $\gamma = 200$. Additionally, $\delta$ (Equation \ref{eq:ar}) was set to 10. We provided detailed information on the model architecture of the proposed Attri-VAE, including our publicly available code, in our GitHub repository \footnote{https://github.com/iremcetin/Attri-VAE}.

The Attri-VAE was trained on an NVIDIA Tesla T4 GPU using Adam optimizer with a learning rate equals to 0.0001 and batch size of 16 for 10000 epochs. The dataset was split into 70/30 training (47 pathological, 23 healthy) and testing (20 pathological and 10 healthy subjects) sets. Subsequently, random oversampling of the normal subjects was employed in the training set as a strategy to treat the unbalanced behavior of the dataset; however, the testing set was kept unchanged. Note that the proposed model is implemented using python programming language and PyTorch library (version 1.10.0) \footnote{https://pytorch.org/}. Image pre-processing and transformations were done using the python-based MONAI library (version 0.8.0) \footnote{https://monai.io/}.

\subsection{Experimental setting and evaluation criteria}
\label{sec:experimental_setting_eval}
The performance of the proposed Attri-VAE, both qualitatively and quantitatively, was compared with VAE, $\beta$-VAE, and AR-VAE. As all three methods are the special cases of the proposed model, we demonstrated this comparison in several ablation studies by removing different components of the proposed network (i.e., $\beta$, MLP, and attribute-regularization (AR) components) such that VAE represents the removal of $\beta$, MLP and AR components of the Attri-VAE, $\beta$-VAE represents Attri-VAE without MLP and AR losses and AR-VAE is Attri-VAE without MLP loss. First of all, the degree of disentanglement of the proposed latent space was evaluated with respect to different data attributes, using the following metrics available in the literature: the modularity metric, to analyse the dependence of each dimension of the latent space on only one attribute \citep{NEURIPS2018_2b24d495Modu} such that if the latent dimension is ideally modular then it will have high mutual information with a single attribute and zero mutual information with others \citep{NEURIPS2018_2b24d495Modu}; the mutual information gap (MIG), to evaluate the MI difference between a given attribute and the top two dimensions of the latent space that share maximum MI with the corresponding attribute \citep{MIG_metric}; the separated attribute predictability (SAP), to measure the difference in the prediction error of the two most predictive dimensions of the latent space for a given attribute \citep{Kumar2018VariationalIO}; and the Spearman correlation coefficient (SCC) score, to compute its maximum value between an attribute and each dimension of the latent space.

In parallel, the interpretability metric introduced in \citep{interp} was used to measure the ability to predict a given attribute using only one dimension of the latent space. For this, a linear probabilistic relationship between the corresponding data attribute and the regularized latent space dimension was calculated. Then, the interpretability score was computed by summing up the logarithms of these probabilities corresponding to each test sample \citep{interp}. As for the $\beta$-VAE model, dimensions having a high MI with the corresponding data attribute were chosen for the interpretability estimation. The reconstruction fidelity performance was also evaluated, employing the maximum mean discrepancy (MMD) score \citep{MMD}, which measures the distance between the distributions of real and reconstructed data examples, as well as their mutual information (MI) as an image similarity metric. The interpretability and MI metrics were then used to identify the optimal values of the most relevant hyperparameters in Equation \ref{loss_all} and Equation \ref{eq:ar} (i.e., $\beta$, $\gamma$, and $\delta$), evaluating the influence of the KL divergence ($\beta$) and attribute regularization ($\gamma$) loss terms, as well as the weight of the distance matrix between two samples in a latent dimension. As a proof-of-concept, the hyperparameter sensitivity analysis was performed with only the four cardiac shape-based interpretable attributes.  

Another set of experiments was carried out to explore the potential of the latent space generated by the Attri-VAE to create synthetically realistic samples. First, two samples in the Attri-VAE latent space, corresponding to input data with distinct cardiac characteristics (e.g., thin vs. thick myocardium, absence vs. presence of myocardium infarct), were chosen as references to synthetically generate interpolated images through their trajectory. Secondly, we qualitatively evaluated the control over individual data attributes during the generation process of the Attri-VAE model. Given a sample with a latent code $z$, a given attribute (e.g., LV volume) can be scanned from low to high values changing the latent code of the corresponding regularized dimension, due to their monotonic relationship. The attribute scanning creates synthetically generated samples in a latent space trajectory where only the chosen attribute is changed, facilitating its interpretation. To further facilitate the identification of each attribute's visual influence in the synthetically generated images, gradient-based attention maps were also estimated. 

Finally, the performance of the Attri-VAE model for classifying healthy and pathological hearts was assessed using the area under the curve (AUC) and accuracy (ACC) metrics, using both the EMIDEC and the ACDC17 challenge datasets. The Attri-VAE results were benchmarked against other VAE-type approaches (VAE+MLP, $\beta$-VAE+MLP), as well as to classical radiomics analysis (with SVM). The latent space projections of the Attri-VAE model, regularized by different attributes, were also qualitatively analysed to identify the attributes better differentiating healthy and pathological clusters of samples. 

\section{Results}
\label{results}

\subsection{Disentanglement and interpretability}
\label{sec:disent_interp}




The proposed Attri-VAE approach obtained better disentanglement metric scores than its ablated variants (i.e., $\beta$-VAE~and AR-VAE) using shape and clinical attributes, implying a more disentangled latent space.


Firstly, all models provided high modularity values (Attri-VAE: 0.98 vs. $\beta$-VAE: 0.97 vs. 0.98 AR-VAE), signaling that each dimension of the latent spaces in all models only depended on one data attribute. The Attri-VAE also resulted in higher MIG/SAP scores than the others (Attri-VAE: 0.60/0.63 vs. $\beta$-VAE: 0.02/0.05 vs. AR-VAE: 0.51/0.55). In its turn, the SCC metric estimated for Attri-VAE was substantially higher than the corresponding $\beta$-VAE one and only slightly higher than the one for AR-VAE (Attri-VAE: 0.97 vs. $\beta$-VAE: 0.46 vs. AR-VAE: 0.96) due to the monotonic relationship between a given attribute and the regularized latent dimension enforced by both Attri-VAE and AR-VAE. When using radiomics features, the same trend was observed, with some Attri-VAE disentanglement metrics (MIG and SAP) slightly lower than when using shape and clinical attributes (Attri-VAE / $\beta$-VAE /AR-VAE): modularity, 0.98/0.98 /0.98; MIG, 0.49/0.01/0.49; SAP, 0.51/0.06/0.51; SCC, 0.98/0.42/0.97). 


\begin{figure*}[h]
\centering
\includegraphics[width=0.9\linewidth]{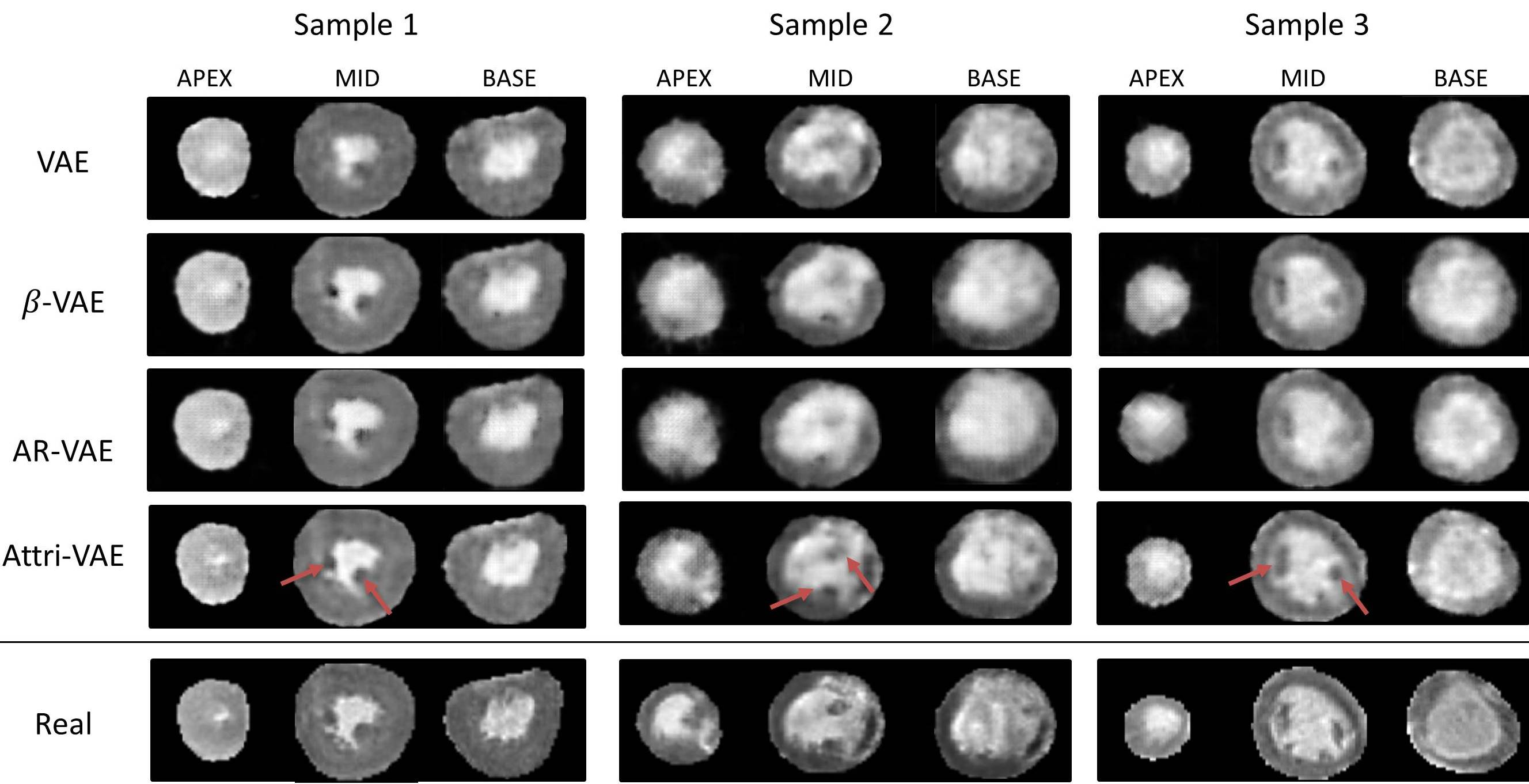}
\caption{Three examples of real and reconstructed images using the VAE, $\beta$-VAE, AR-VAE, and Attri-VAE approaches. Three slices are shown in every example: apical (APEX), mid-ventricle (MID), and basal (BASE) slices. Sample 1 and 3 correspond to healthy hearts while Sample 2 shows an infarcted myocardium.}
\label{fig:reconstructions}
\end{figure*}

\begin{table}[h]
\small
\centering
\caption{Interpretability score \citep{interp} of most relevant shape, clinical and radiomics attributes, as encoded in the latent space, with the Attri-VAE, $\beta$-VAE, and AR-VAE approaches. LV: left ventricle, MYO: myocardium, EF: ejection fraction. Max: maximum, DE: difference entropy, ZP: zone percentage. Maximum interpretability is 1.0.}
\label{tab:interpretability_vals}
\begin{tabular}[t]{|l|c|c|c|}
\hline
&Attri-VAE & $\beta$-VAE &  AR-VAE\\
\hline 
LV volume &0.89 &0.14 &  0.80\\
MYO volume & 0.93 &0.02 &  0.87\\
Wall thickness  &0.95 & 0.10 &  0.90 \\
EF  & 0.94 & 0.03 &  0.90\\
Gender  & 0.98 & 0.19 &  0.94\\
Age  & 0.93 & 0.12 &  0.84\\
Tobacco  & 0.70 & 0.19 & 0.74\\
Radiomics & 0.91 & 0.06 & 0.90\\
\hline
\end{tabular}
\end{table}%

Table \ref{tab:interpretability_vals} shows the interpretability scores for Attri-VAE, $\beta$-VAE, and AR-VAE obtained with clinical and shape descriptors as well as radiomics features. The radiomics feature selection identified seven of them having the most discriminative power: four shape-based, being the sphericity of the left ventricle, the maximum 2D diameter of the myocardium, as well as left ventricle and myocardial volumes; three texture-based, being the correlation of the left ventricle, the difference entropy of the myocardium and the inverse variance of the left ventricle. 

It can easily be observed that the Attri-VAE provided a high degree of interpretability (i.e., close to 1.0) for all attributes, with the exception of tobacco (0.70). Among shape and clinical features, gender was the attribute with a higher interpretability (0.98), followed by the wall thickness (0.95), meaning that they could be predicted with only one dimension of the latent space. As for radiomics features, the average interpretability metric value was 0.91, with shape-based ones showing slightly larger values than texture features (0.93 and 0.89, respectively); the maximum 2D diameter of the myocardium presented the highest value (0.97). On the other hand, the $\beta$-VAE clearly resulted in lower interpretability values (average of 0.11 for shape/clinical attributes and 0.06 for radiomics features)  and AR-VAE obtained slightly lower interpretability metric score (average of 0.86 for shape/clinical attributes and 0.90 for radiomics features) than the proposed Attri-VAE.







\subsection{Reconstruction fidelity}

 Reconstruction fidelity is another important factor as the high degree of disentanglement and interpretability should not be accompanied by a reduced reconstruction. Hence in this experiment, we demonstrated the reconstruction quality of the proposed approach by comparing its performance with its ablated variants, specifically VAE, $\beta$-VAE, and AR-VAE.

 The proposed Attri-VAE approach obtained the lowest MMD values, representing a lower distance between input and reconstructed images (see Table \ref{tab:recons_metrics}). We observed that removing only the MLP loss from Attri-VAE (AR-VAE) obtained the best MI value of 0.92 (0.91 and 0.89 for VAE and Attri-VAE, respectively). 

\begin{table}[h]
\centering
\caption{Ablation study on the reconstruction accuracy of Attri-VAE on the EMIDEC dataset, quantified with the maximum mean discrepancy (MMD) and mutual information (MI) metrics. The MMD results are given as $\pm$ standard deviation. w/o : without, AR: attribute-regularization}
\label{tab:recons_metrics}
\begin{tabular}[t]{|l|c|c|}
\hline
&MMD $\times 10^{2}$ &MI\\
\hline
VAE (w/o $\beta$, MLP and AR) &1.86 $\pm$ 0.06 & 0.91\\
$\beta$-VAE (w/o MLP and AR) &1.38 $\pm$ 0.04 & 0.87\\
AR-VAE (w/o MLP) & 1.74 $\pm$ 0.06 &  0.92 \\
Attri-VAE & 1.18 $\pm$ 0.03 & 0.89 \\
\hline
\end{tabular}
\end{table}%

The reconstructions of three data examples from the EMIDEC dataset where the performance of the Attri-VAE was compared with its different variants, can be seen in Figure \ref{fig:reconstructions}. Even though all models achieved similar qualitative reconstruction results, the Attri-VAE model generated images better preserving the heart shape and details than the other models: see the papillary muscles in mid-myocardium slices (dark regions in the blood pool) or the left ventricular cavity in apical slices of Sample 2 and Sample 3 in Figure \ref{fig:reconstructions}. We can also observe in the figure that apical slices were more difficult to reconstruct than mid-ventricle and basal ones for the three tested models. 

\subsection{Hyperparameter sensitivity analysis}
\begin{figure}[h]
\centering
\includegraphics[width=8.1cm]{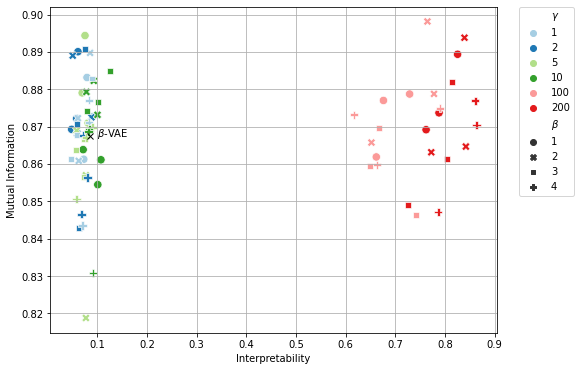}
    \caption{Effect of hyperparameters on the interpretability and reconstruction fidelity of the Attri-VAE approach. The hyperparameters $\beta$ and $\gamma$ of the Attri-VAE model control the influence of the loss terms for the Kullback-Leibler divergence between learned prior and posterior distributions, and attribute regularization, respectively. Each marker represents a unique combination of the hyperparameters $\beta$ and $\gamma$, which is indicated by color and shape, respectively.  For comparison, the performance of $\beta$-VAE ($\beta$ = 3) is also represented. Best performance combinations are located in the top right corner of the graph.}
\label{fig:hyperparam_experiment}
\end{figure}
This study evaluates the impact of hyperparameters on the interpretability and reconstruction quality of the Attri-VAE. The trade-off between reconstruction quality and interpretability can be seen in Figure \ref{fig:hyperparam_experiment} where the performance of $\beta$-VAE ($\beta$ = 3) is also provided for comparison. A visual inspection of the figure suggests that $\gamma$, i.e., the hyperparameter controlling the attribute regularization, was the key to obtaining good interpretability values while keeping reasonable reconstruction fidelity (mutual information $\geq$ 0.88), with values of $\gamma \geq$ 100.

On the other hand, the $\beta$ hyperparameter was not as relevant as the $\gamma$ hyperparameter. As expected, the $\beta$-VAE provided acceptable reconstruction fidelity results but low values of interpretability. We need to point out that the same results were obtained when using radiomics features instead of shape-based attributes.

\subsection{Latent space interpolation and attribute scanning}

This experiment aims to qualitatively evaluate the proposed latent space by demonstrating its ability to interpolate between different data examples and control individual attributes during the generation process.\\
\begin{figure}[ht]
\centering
\includegraphics[width=8.2cm]{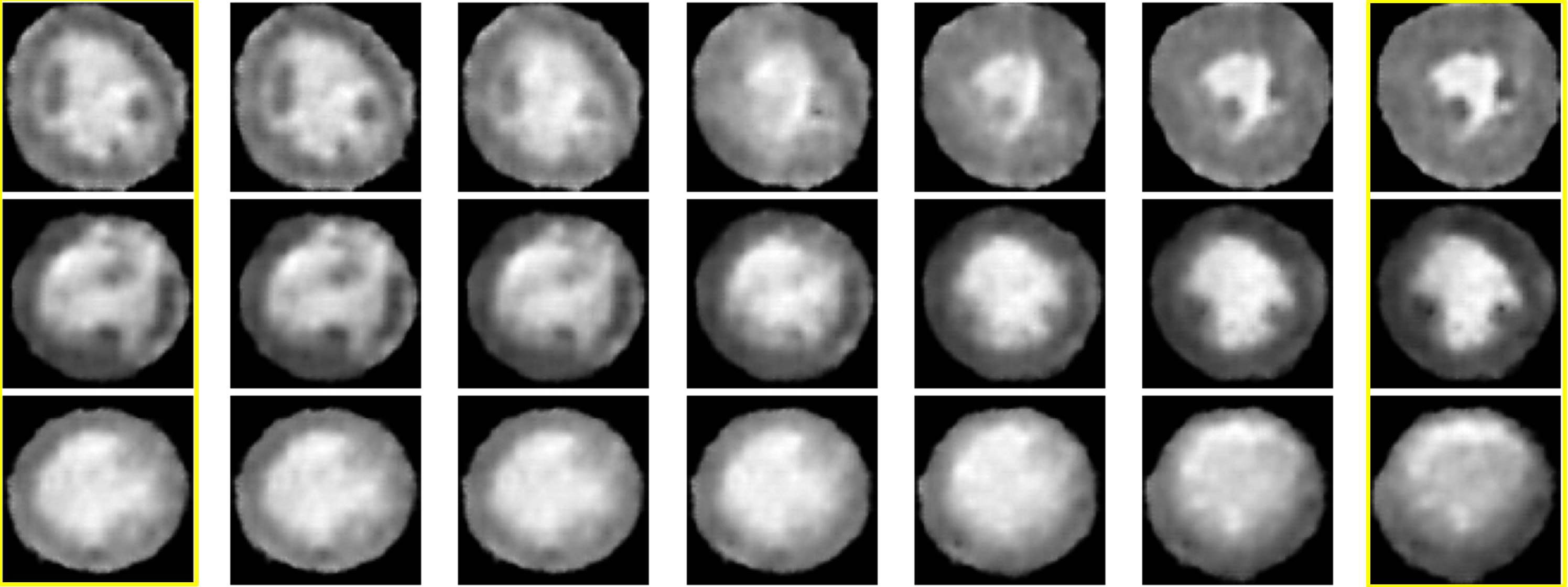}
\caption{Linear latent space interpolation between two data samples (extremes of each row in yellow frames) from the EMIDEC dataset. Each row depicts the interpolation from the left to the right latent vector dimension. Top: from thin to thick myocardium. Middle: from a myocardium with a scar to one without. Bottom: from a healthy subject to a patient with a myocardial infarct.}
\label{fig:interpolation}
\end{figure}
\begin{figure*}[ht]
\centering
\includegraphics[width=16cm]{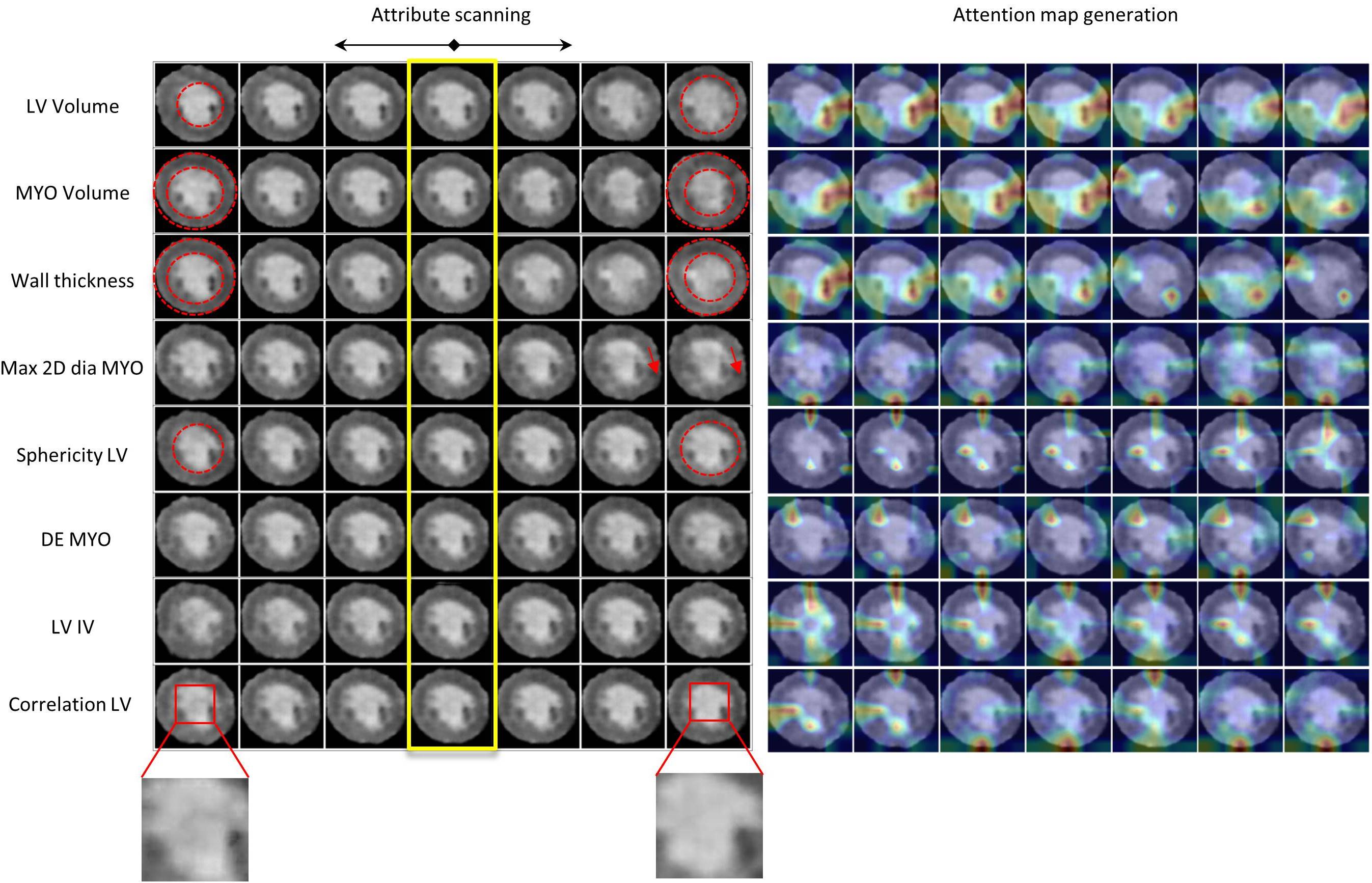}
\caption{Scanning of attributes and corresponding gradient-based attention maps for shape and radiomics features. The image in the middle (4th column, in yellow frame) shows the original reconstructed image. DE: difference entropy, IV: inverse variance, Max 2D dia: maximum 2-dimensional diameter, LV: left-ventricle, MYO: myocardium. Note that the first three rows demonstrate the attribute scanning that was done on the latent space of Attri-VAE, which was trained with clinical and shape features. The remaining rows represent the attribute scanning on the latent space of Attri-VAE trained with selected radiomics features.}
\label{fig:manipulation_interpretable}
\end{figure*}
For the first experiment, the interpolation was employed, between distinct and well-separated samples in the learned latent space of the Attri-VAE. The proposed approach generates synthetic interpolated images that have a realistic appearance, gradually changing the main sample characteristics in the trajectory between the chosen samples. The first row of Figure \ref{fig:interpolation} clearly demonstrates the Attri-VAE model's ability to create smooth transitions between hearts having largely different characteristics such as (thin to thick) wall thickness. The other two rows of the figure demonstrate a similar behavior from non-infarcted/scar to infarcted/scar patients.

The second experiment presents the effect of scanning an individual attribute along its corresponding regularized dimension in the Attri-VAE model, where all the remaining attributes remain fixed. The first three rows of Figure \ref{fig:manipulation_interpretable} exemplify the attribute scanning that was done on the latent space of Attri-VAE, which was trained with clinical plus shape features. The rest of the rows represent the attribute scanning on the latent space of Attri-VAE trained with selected radiomics features. For shape-based attributes, the changes in the attribute when moving along different values of the regularized dimension are clearly seen. For instance, from the left to the right in Figure \ref{fig:manipulation_interpretable}, how LV and myocardial volumes are increasing in the first and second rows, respectively, or how the LV becomes more spherical. More subtle changes are observed with texture-based radiomics but they can still be identified with a careful inspection of the generated images. For example, moving along the latent space dimension corresponding to the correlation LV, we find more or less intensity homogeneity in the LV. The LV inverse variance (LV-IV) and the difference entropy of the myocardium (DE-MYO) only produced small changes that consisted in slightly thicker myocardium with lower values of LV-IV (left samples in Figure \ref{fig:manipulation_interpretable}) and some darker patches and heterogeneous texture in the myocardium for higher values of DE-MYO (right in Figure \ref{fig:manipulation_interpretable}). It needs to be pointed out that attribute scanning for clinical attributes such as age, gender, and tobacco is not shown since the images do not visually change along the corresponding regularized dimensions.

To further illustrate the impact of each studied attribute, we have also generated the gradient-based attention maps linked to the changes in each regularized latent dimension of the Attri-VAE model. Attention maps show the high response regions in the images when changing specific dimensions of the latent space that correspond to specific attributes. We can see in Figure \ref{fig:manipulation_interpretable} that more attention (i.e., higher response) is paid to more varying regions for shape-based attributes (e.g., right side of the slide for LV volume, where LV is increasing from the left to the right in the regularized dimension). In general, attention maps for texture-based features have less high-response regions than for shape-based attributes. However, in some texture-based features such as the difference entropy of the myocardium, a higher response can still be localized (in this example, darker regions in the top left part of the slice). On the other hand, interpretation and validation of the resulting attention maps for other attributes such as for LV-IV are more challenging. 
\subsection{Classification}
This experiment illustrates the classification performance of the proposed Attri-VAE. Additionally, we have also compared its performance with several models, such as radiomics analysis and its ablated versions. The best result was obtained in both EMIDEC and ACDC datasets with the Attri-VAE trained with radiomics features (accuracy of 0.97 and 0.58 for both datasets), while radiomics+MLP was the worst for EMIDEC (accuracy of 0.76) and the $\beta$-VAE+MLP for ACDC (accuracy of 0.45) (see Table \ref{tab:prediction_results}). 
There were only minor differences in the accuracy of the Attri-VAE method when trained with clinical and shape attributes or radiomics features. All the evaluated models, trained with the EMIDEC data, substantially dropped their performance when tested on the external ACDC dataset, especially the VAE-based approaches.
\begin{table}[h]
\centering
\caption{Ablation study on the classification performance of Attri-VAE with EMIDEC and ACDC datasets (healthy vs. myocardial infarction) with different models. The results are reported as accuracy / AUC scores. SVM: support vector machine, w/o : without, AR: attribute-regularization}
\label{tab:prediction_results}
\begin{tabular}[t]{|l|l|c|}
\hline
&EMIDEC &ACDC\\
\hline
Attri-VAE (Clinical+Shape) & 0.93 / 0.94 & 0.55 / 0.54\\
Attri-VAE (Radiomics) & 0.97 / 0.96  & 0.58 / 0.52 \\
$\beta$-VAE+MLP (w/o AR) & 0.90 / 0.90 & 0.45 / 0.31\\
VAE+MLP (w/o $\beta$ and AR) & 0.87 / 0.80 & 0.53 / 0.35\\
Radiomics analysis (SVM) & 0.77 / 0.75 & 0.60 / 0.61 \\
Radiomics analysis (MLP) & 0.76 / 0.75  & 0.58 / 0.60  \\
\hline
\end{tabular}
\end{table}%
\begin{figure}[t]
\centering
\includegraphics[width=8.3cm]{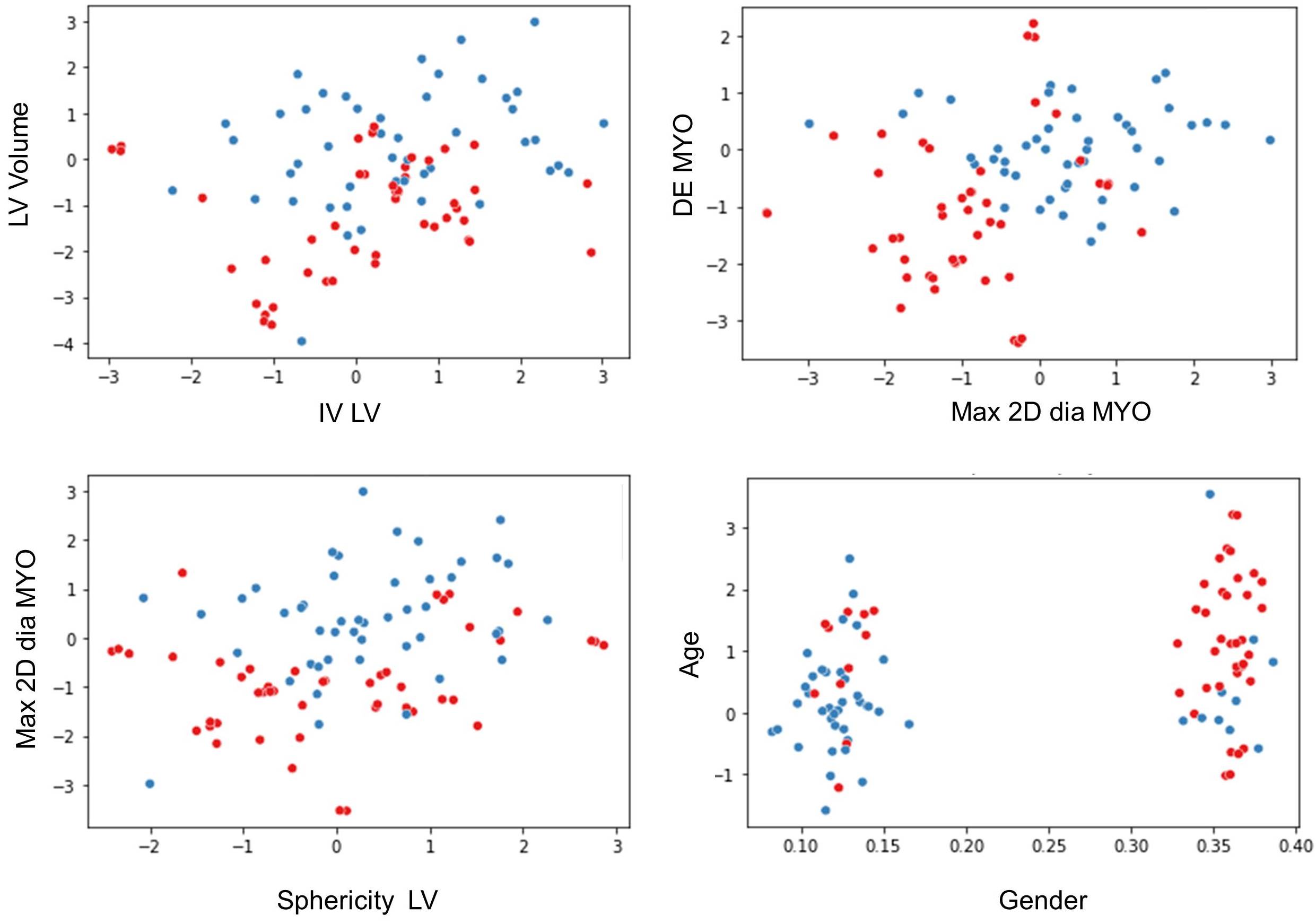}
\caption{Latent space projections of regularized dimensions for different clinical, shape, and radiomics attributes. Each point in the graphs represents a healthy or a myocardial infarction patient (red and blue, respectively), LV: left-ventricle, MYO: myocardium, IV: inverse variance, DE: difference entropy, Max 2D dia: maximum 2-dimensional diameter.}
\label{fig:prediction_latentspaceplots}
\end{figure}

Finally, the latent space projections of different regularized latent dimensions are visualized in Figure \ref{fig:prediction_latentspaceplots}, with plot axes representing the encoded data attributes. As it can be observed in the figure, our model is able to build several reduced dimensionality spaces, based on different attributes, where healthy and pathological cases (red and blue in the figure, respectively) can easily be clustered. For instance, the maximum 2D diameter of the myocardium and the LV volume attributes correctly separate most samples into two clusters. Interestingly, despite Attri-VAE having poor control over clinical attributes such as age or gender, they also facilitate the construction of the latent spaces and sample discrimination, as can be seen in the gender-age plot of Figure \ref{fig:prediction_latentspaceplots}. 

\section{Discussion}
\label{discussion}
The analysis of medical data demands for interpretable methods. However, the majority of deep learning methods do not fulfill the minimum level of interpretability to be used in reasoning medical decisions \citep{sergio19}, being difficult to relate clinically and physiologically meaningful attributes with model parameters and outcomes. Fortunately, interpretable and explainable deep learning methods are starting to emerge. Models creating latent space representations, such as variational autoencoders, are promising but attributes are usually entangled in the resulting reduced dimensionality space, hampering its interpretation. In this work, we have presented the Attri-VAE approach that generates disentangled and interpretable representations where different types of attributes (e.g., clinical, shape, radiomics) are individually encoded into a given dimension of the resulting latent space. 
The results obtained by the proposed Attri-VAE model based on disentanglement and interpretability metrics clearly outperformed the state-of-the-art $\beta$-VAE approach and obtained slightly better results than AR-VAE, indicating a high degree of disentanglement and a monotonic relationship between a given attribute and the corresponding regularized dimension. However, Attri-VAE values for some metrics such as the MIG and SAP, although substantially better than those of $\beta$-VAE, were far from the maximum (e.g., 1.0). The same trend was observed by \citep{arvae} in the MNIST (i.e., for digit number identification) dataset, suggesting that other latent dimensions, beyond the regularized ones, share a high MI with different attributes. We would like to point out here that the performance of Attri-VAE was compared with several methods including, AR-VAE \citep{arvae}, further work is needed to compare the performance of the proposed approach with additional models, such as Guided-VAE \citep{ding2020guided}. More work is also required to visually evaluate the model interpretability by human experts in addition to the quantitative results, in order to better illustrate the interpretation performance of the proposed approach.

Hyperparameter selection was a key step for finding the optimal Attri-VAE configuration providing an excellent trade-off between reconstruction fidelity, at the level of state-of-the-art alternatives, and interpretability; even though the Attri-VAE approach had a more constrained latent space, it generated reconstructions that are less smooth than other VAE-based models and more similar to the original input images. This explains why the proposed Attri-VAE had a slightly lower MI value and higher MMD and interpretability values than the other models. The most critical parameter to enforce interpretability was the weight of the attribute regularization loss term ($\gamma$ in Equation \ref{loss_all}).
The Attri-VAE plot of reconstruction fidelity vs. interpretability, shown in Figure \ref{fig:hyperparam_experiment} had the same pattern as the one obtained by \citep{arvae}. Interestingly, their optimal $\gamma$ values were lower than ours ([5.0, 10.0] vs $\geq$ 100), likely due to the higher complexity of the cardiac MRI data and corresponding latent space compared to the MNIST dataset. On the other hand, the best $\delta$ values were the same in the two studies ([1.0, 10.0]). However, it is worth noting that we did not evaluate the effect of the size of the latent space and the position of regularized dimensions. Therefore, future work is needed to analyze the size of the latent space and the importance of the position of regularized dimensions by, for example, randomly replacing them during inference. Additionally, more work is required to determine how various losses (i.e., classification and attribute losses) affect the latent space as a whole and if this is dependent on the latent space's dimensionality.

One of the most interesting characteristics of the Attri-VAE approach is the ability to create realistic synthetic data by sampling the created latent space and interpolating between different original reconstructed inputs, which can be very useful for controllable and attribute-based data augmentation of training datasets in machine learning applications. Scanning a regularized dimension of the latent space creates synthetic images where the corresponding attribute changes its values, as can easily be observed for shape descriptors (e.g., LV and myocardial volumes, wall thickness) in Figure \ref{fig:manipulation_interpretable}. In addition, the proposed approach allows a better understanding of some (texture-based) radiomics features, which are often difficult to interpret. However, clinical attributes such as age, gender, or tobacco consumption, despite obtaining good interpretability scores, did not create visually different interpolated samples over the regularized dimensions. One potential reason is the difficulty of the attribute regularization to control binary attributes, as suggested by \citep{arvae}. Furthermore, the studied clinical attributes cannot be disassociated from the shape and image intensity variations (e.g., morphological changes of the heart with age), thus it is too restrictive to keep all attributes fixed except a clinical one. In consequence, we have employed the clinical attribute referring to healthy vs. myocardial infarction as a task-specific label, as the aim of this classification is to be able to separate different groups of patients in the latent space, by enforcing the continuous variables to be able to predict the desired class and provide interpretable results. More work is needed to better construct latent spaces where clinical information can be disentangled from other attributes. Additionally, we would like to point out that the proposed Attri-VAE was trained on delineated left ventricle images; future work needs to include other cardiac structures to integrate global changes in cardiac tissue and shape.

The generated gradient-based attention maps contributed to locally identifying the cardiac regions where the attributes were influencing, which was particularly useful for global attributes and for complex features such as the texture ones. Additionally, we only employed the well-known Grad-CAM method, which could be complemented with additional interpretability methods (e.g., LIME and its variations \citep{LIME_Ribeiro0G16}) to better understand the attribute effects on the latent space. However, the attention maps have some limitations that have already been addressed in different studies in the literature, which demonstrate that saliency maps underperform in some key criteria such as localization, parameter sensitivity, repeatability, and reproducibility \citep{Arun_atten_map}. Thus the reliability of attention maps still requires further investigation to assess its robustness and reliability with respect to data input and model parameter perturbations \citep{pmid32510054}. In parallel, enhanced 3D visualizations of the generated samples are needed to have an overall perspective of the cardiac differences, beyond 2D slice views of the resulting images. 

The proposed Attri-VAE model also achieved excellent classification performance (healthy vs. myocardial infarction), outperforming the other VAE-based approaches, with slightly better results when trained with radiomics. When evaluated with the EMIDEC training dataset using ground-truth labels, the Attri-VAE approach provided accuracy results (0.98) equivalent to the best challenge participants reporting their performance on the same dataset (1.0 \citep{lourenco}, 0.95 \citep{shi}, 0.94 \citep{ivantsits} and 0.90 \citep{sharma}). For the testing EMIDEC dataset \citep{Lalande2021DeepLM}, the best participant method obtained a decreased accuracy (0.82, \citep{lourenco, girum}), increasing to 0.92 for the challenge organizers \citep{shi}. As for the ACDC dataset, which was tested as an external database (i.e., without considering it in training), classification accuracy was substantially reduced (0.58), being worst than results reported by challenge participants \citep{bernardTMI} (0.96) to classify between the different pathologies (not only between healthy and myocardial infarction). On the other hand, we would like to point out that these two datasets have employed different image acquisition techniques and contain images from different imaging modalities, such that the EMIDEC dataset consists of DE-MRI images and the ACDC dataset contains cine-MRI images. However, further work is required to find out the main reason behind this performance drop including evaluating the Attri-VAE's performance in comparison with an additional model, such as a baseline CNN. Additionally, more work is also needed to improve the reconstruction quality and the generalization of the Attri-VAE model to unseen data, being more robust to different quality and imaging acquisition protocols, through domain adaptation techniques or image registration or integrating these differences into its latent space, using databases such as the M\&Ms challenge \citep{9458279}.

One limitation of the Attri-VAE approach, also acknowledged by Pati and Lerch \citep{arvae}, is the dependence on the selection of the data attributes to train the model. An incorrect attribute selection could lead to undesired strong correlations of several attributes that will not ensure a monotonic relationship with the corresponding regularized dimension, leading to less attribute interpretation and reconstruction quality. However, the projection of original samples in latent spaces with regularized dimensions for different attributes (see Figure \ref{fig:prediction_latentspaceplots}) could be used as an interpretable attribute selection, identifying the ones better separating the analyzed classes such as the maximum 2D diameter of the myocardium and the LV volume attributes in our experiments. Further work will focus on fully integrating advanced feature selection techniques with the Attri-VAE model, as well as exploring alternative interpretability methods (see the recent review of Salahuddin et al. \citep{SALAHUDDIN2022105111}) to better understand the role of clinical and imaging attributes on medical decisions in cardiovascular applications. Self-supervision will also be explored, as an opportunity to make use of unlabeled data to further improve our results.


%

\section{Conclusions}
\label{conclusions}
We have presented an approach, referred to as Attri-VAE, which implements attribute-based regularization in a $\beta$-VAE scheme with a classification module for the purpose of attribute-specific interpretation, synthetic data generation, and classification of cardiovascular images. The basis of the proposed Attri-VAE model is to structure its latent space for encoding individual data attributes to specific latent dimensions, being guided by an attribute regularization loss term. The resulting constrained latent space can be easily manipulated along its regularized dimensions for an enhanced interpretation of different attributes. Additionally, the proposed approach improves the current state-of-the-art for classifying cardiovascular images and allows the visualization of the most discriminative attributes by projecting the trained latent space. Future work will be focused on improving the generalization of the trained Attri-VAE model to images with different acquisition characteristics. 

\section{Competing interests}
The authors declare that they have no competing interests.
\section{Author contributions}
All authors participated in the critical revision of the manuscript, and final approval of the submitted manuscript. IC, MAGB, and OC contributed to the study concepts, methods, and underlying data collection. IC, MS, MAGB, and OC drafted the manuscript. IC, MAGB, and OC designed the machine learning methods. IC performed the data pre-processing and data analysis.
\section{Funding}
This work was partly funded by the European Union’s Horizon 2020 research and innovation programme under grant agreement No 825903 (euCanSHare project). 
\section{Availability of data and materials}
This research was conducted using the publicly available EMIDEC and ACDC datasets. These datasets can be accessed in \url{http://emidec.com/dataset} and \url{https://www.creatis.insa-lyon.fr/Challenge/acdc/}.
We have also made our code publicly available and can be found in \url{https://github.com/iremcetin/Attri-VAE}
\section{Ethical approval}
The datasets employed in this study are publicly available sources. Thus, for the detailed information see \url{http://emidec.com/} and \url{https://www.creatis.insa-lyon.fr/Challenge/acdc/}. 
\bibliographystyle{model2-names}\biboptions{authoryear}
\bibliography{mybibfile}

\end{document}